\documentclass[aps,twocolumn]{revtex4-1}
\usepackage{graphicx}
\usepackage{amsmath}
\usepackage{color}
\usepackage{bm}

\newcommand{\mI}{\mathcal{I}}
\newcommand{\mN}{\mathcal{N}}
\newcommand{\mD}{\mathcal{D}}
\newcommand{\mH}{\mathcal{H}}
\newcommand{\mZ}{\mathcal{Z}}
\newcommand{\gcol}[1]{{\color{black} #1}}

\begin{document}

\title{Statistical Features of High-Dimensional Hamiltonian Systems}

\author{Marco Baldovin}
\affiliation{Dipartimento di Fisica, Universit\`a di Roma ``Sapienza'', Piazzale A. Moro 2, I-00185, Roma, Italy}

\author{Giacomo Gradenigo}
\affiliation{Gran Sasso Science Institute, Viale F. Crispi 7, 67100 L'Aquila, Italy}
\affiliation{INFN-Laboratori Nazionali del Gran Sasso, Via G. Acitelli 22, 67100 Assergi (AQ), Italy}

\author{Angelo Vulpiani}
\affiliation{Dipartimento di Fisica, Universit\`a di Roma ``Sapienza'', Piazzale A. Moro 2, I-00185, Roma, Italy}

\begin{abstract}
In this short review we propose a critical assessment of the role of
chaos for the thermalization of Hamiltonian systems with high
dimensionality. We discuss this problem for both classical and quantum
systems. A comparison is made between the two situations: some
examples from recent and past literature are presented which support
the point of view that chaos is \emph{not} necessary for
thermalization. Finally, we suggest that a close analogy holds between
the role played by Kinchin's theorem for high-dimensional classical
systems and the role played by Von Neumann's theorem for many-body quantum
systems.
\end{abstract}

\maketitle

\tableofcontents

\section{Introduction}
\label{sec:intro}

Hamiltonian systems have a great relevance in many fields, from
celestial mechanics to fluids and plasma physics, not even to mention
the importance of the Hamiltonian formalism for quantum mechanics and
quantum field theory~\cite{book-TDLee81}. The aim of this contribution
is to discuss the relation between the dynamics of Hamiltonian systems
with many degrees of freedom and the foundations of statistical
mechanics~\cite{book-CFLV08,book-CCV09,book-O13}. In particular we
want to address the problem of thermalization in integrable and
nearly-integrable systems, questioning the role played by chaos in
these processes. The problem of thermalization of high-dimensional
integrable Hamiltonian systems is, in our opinion, a truly
relevant one and is nowadays of groundbreaking importance for one
main reason: it has a direct connection with the problem of
thermalization of quantum systems. If the problem of thermalization in
integrable Hamiltonian systems is analogous to that of thermalization
in quantum mechanics, it is then possible to take advantage of the
rich conceptual and methodological framework developed for classical
systems to investigate properties of the quantum ones. The point of
view that we want to promote here is that the presence or the absence of
thermal equilibrium is not an \emph{absolute} property of the system
but it depends on the set of variables chosen to represent it.
We want to stress that, in spite of the role of the used variables,
this is not a subjective point of view. We
will present some examples supporting the idea that the notion of
thermal state strongly depends on the choice of the variables used to
describe the system. It can be asked at this point why the study of
thermalization should still be an interesting problem. The point is
that the most recent technologies allow us for manipulation of
mesoscopic and quantum systems in such a way that foundational
questions which have been experimentally out-of-reach so far and are
now being considered with renewed interest. Not last, the storage and
erase of information due to the (lack of) thermalization is of
paramount importance for all applications to quantum computing.\\

  The discussion develops through sections as follows: in
  Sec.~\ref{sec:s1} the standard viewpoint, according to which chaos
  plays a key role to guarantee thermalization, is summarized ; in
  Sec.~\ref{sec:s2} we present three counter-examples to the above
  view and provide a more refined characterization of the central
  point in the ``thermalization problem'', namely the choice of the
  variables used to describe the system; in Sec.~\ref{sec:s3} we
  discuss the analogies between the problem of thermalization in
  classical and quantum mechanics and we compare the scenarios
  established respectively by the ergodic theorem by Khinchin for
  classical systems~\cite{book-K49} and the quantum ergodic theorem by
  Von Neumann~\cite{JVN10,GLMTZ10,GLTZ10}, accounting for open
  questions and future perspectives in the field.

\section{The role of KAM and  chaos  for the  timescales}
\label{sec:s1}

It is usually believed that, in order to observe thermal behaviour in
a given Hamiltonian system, ergodic hypothesis has to be
satisfied~\cite{book-O13}. Let us just recall that such hypothesis,
first introduced by Boltzmann, roughly speaking states that every
constant-energy hypersurface in the phase space of a Hamiltonian
system is totally accessible to any motion with the corresponding
energy. In modern terms, the assumption of ergodicity amounts to ask
that phase space cannot be partitioned in disjoint components which
are invariant under time
evolution~\cite{book-CFLV08,book-EL02}. Indeed, if the dynamics can
explore the whole hypersurface, the latter cannot be divided into
measurable regions invariant under time evolution without resulting in
a contradiction.  The existence of conserved quantities different from
energy implies in turn the existence of other invariant hypersurfaces,
which prevent the system to be ergodic.

The claim that thermalization is related to ergodicity summarizes the
viewpoint of those who think that the presence of \textit{invariant
  manifolds} in phase space, which work as \emph{obstructions} to
ergodicity, is the main obstacle to thermal behaviour.  The problem of
\emph{``thermalization''} can be thus reformulated as the
determination of the existence of integrals of motion other than
energy, solved by Poincar\'e in his celebrated work on the three-body
problem~\cite{book-CFLV08,book-CCV09}.  In a few words: consider an
almost integrable system whose Hamiltonian can be written as
\begin{align}
\label{eq:ham}
H=H_0({\bf I}) + \epsilon H_I({\bf I}, {\bf \theta}) \, , 
\end{align}
where $H_0$ only depends on the ``actions'' $\{I_n\}$, with
$n=1,...,N$, and it is independent of the corresponding ``angles''
$\{\theta_n\}$. In the above equation $\varepsilon$ is a (small)
constant: in absence of perturbations, i.e.  $\epsilon=0$, there are
$N$ conservation laws and the motion takes place on $N$-dimensional
invariant tori. Poincar\'e showed that, for $\epsilon \neq 0$,
excluding specific cases, there are no conservation laws apart from
the trivial ones.  In 1923 Fermi generalized Poincar\'e's theorem,
showing that, if $N >2$, is not possible to have a foliation of phase
space in invariant surfaces of dimension $2N - 2$ embedded in the
constant-energy hypersurface of dimension $2N - 1$. From this result
Fermi argued that generic Hamiltonian systems are ergodic as soon as
$\epsilon \neq 0$. Assuming the perspective that global invariant
manifolds are the main obstruction to ergodization and the appearance
of a thermal behaviour, from the moment when Fermi furnished a further
mathematical evidence that such regular regions were absent in
non-integrable systems the expectation of good thermal behaviour for
any such system was quite strong.
It was then a numerical experiment realized by Fermi himself in
collaboration with J. Pasta, S.Ulam and M. Tsingou (who did not appear
among the authors) which showed that this was not the
case~\cite{book-G08}.
Fermi and coworkers studied a chain of weakly nonlinear oscillators,
finding that the system, despite being non-integrable, was not showing
relaxation to a thermal state within reasonable times when initialized
in a very atypical condition. This observation raised the problem that
while the absence of globally conserved quantities guarantees that no
partitioning of phase space takes place, on the other hand it does not
tell anything on the \emph{time-scales} to reach
equilibrium~\cite{BCP13}.  A first understanding of the slow FPUT
time-scales from the perspective of phase-space geometry came from the
celebrated Kolmogorov-Arnold-Moser (KAM) theorem, sketched by
A.N. Kolmogorov already in 1954 and completed
later~\cite{book-CFLV08,book-CCV09}. The theorem says that for each
value of $\epsilon \neq 0$, even very small, some tori of the
unperturbed system, the so-called resonant ones, are completely
destroyed, and this prevents the existence of analytical integrals of
motion.  Despite this, if $\epsilon $ is small, most tori, slightly
deformed, survive; thus the perturbed system (for ``non-pathological''
initial conditions) has a behaviour quite similar to an integrable
one.\\ \\

\subsection{KAM scenario, chaos and slow timescales}

After the discovery of the KAM theory the state-of-the-art for the
dynamical justification of statistical mechanics problem was as
follows: also in weakly non-linear systems phase space is
characterized by the presence of invariant manifolds which, even if
not partitioning it in disjoint components, might crucially slow down
the dynamics. The foundational problem thus turn from \emph{``do we
  have a partitioning of phase space''} to \emph{``how long does it
  takes in the large-$N$ limit?''}.\\

A crucial role in the justification of thermal properties in classical
Hamiltonian system was then played by the notion of chaos. Very
roughly, chaos means exponential divergence of initially nearby
trajectories in phase space. The rate of divergence
$\tau_{\text{div}}$ of nearby orbits is the inverse of a quantity
known as the first (maximal) Lyapunov exponent~\cite{book-PP16}
$\lambda_1$: $\tau_{\text{div}} = 1/\lambda_1$.  Of course the
behaviour of nearby trajectories may, and actually does, depend on the
phase space region they start from. In fact for a system with $N$
degrees of freedom one finds $N$ independent Lyapunov exponents: in
the multiplicity of Lyapunov exponents is encoded the property that
the rate of divergence of nearby orbits has fluctuations in phase
space. This is indeed the picture which comes from both the KAM
theorem and the evidence of many convincing numerical experiments: at
variance with dissipative systems (characterized by a sharp threshold
from regular to chaotic behavior) Hamiltonian systems are
characterized at any energy scale by the coexistence of regular and
chaotic trajectories. In particular one has in mind that fast
ergodization takes place in chaotic regions while the slow timescales
are due to the slow intra-region diffusion. The slow process is thus
controlled by the location and structure of invariant manifolds which
survive in the presence of nonlinearity. The whole matter of
determining on which time-scale the system exhibits thermal properties
boils down to estimating the time-scale of diffusion across chaotic
regions in phase space: this is the so-called Arnold's diffusion.\\ \\

\subsection{Arnold's diffusion}

Arnold's diffusion is the name attached to the slow diffusion taking
place in a phase space characterized by the coexistence of chaotic
regions and regular ones~\cite{book-CFLV08,book-CCV09}. Typically, chaotic regions in
high-dimensional phase spaces do not have a smooth geometry and are
intertwined in a very intricate manner with invariant
manifolds. Diffusion across the whole phase space is driven by chaotic
regions, but their complicated geometry, typically fractal, slows down
diffusion at the same time~\cite{KB85}. From this perspective, let us try to take
a look to the scenario for diffusion in phase space in the case of
weakly non-integrable systems in the limit of large $N$~\cite{FMV91,YK20,HGO94}. Is it
compatible with the appearance of a thermal state?\\

Let us consider the generic Hamiltonian for a weakly-nonintegral
system. This is a system with $N$ degrees of freedom where, after
having introduced the Liouville-Arnold theorem's action-angle variables
$\theta_n(t),\mI_n(t)$, the symplectic dynamics is defined as follows:
\begin{align}
\label{eq:hamper}
\theta_n(t+1) &= \theta_n(t) +\mI_n(t) \,\,\,  , \,\,\, mod \,\, 2 \pi \\
\mI_n(t+1) &= \mI_n(t) + \varepsilon~{ \partial  F(\theta (t+1)) \over \partial \theta_n(t+1) } 
 \,\,\,  , \,\,\, mod \,\, 2 \pi
\end{align}
where $n=1, ... , N$, $F(\theta )$ is the term which breaks
integrability and $\varepsilon$ is an adimensional coefficient which
we assume to be small. Let us note that a symplectic map with $N$
degrees of freedom can be seen as the Poincar\'e section of a
Hamiltonian system with $N+1$ degrees of freedom. The key point for the
purpose of our discussion is to understand what happens to invariant
manifolds in the large-$N$ limit. Several numerical works, see
e.g.~\cite{FMV91}, showed that the volume of regular regions decreases
very fast at large $N$.  In particular the normalized measure
$P(N,\epsilon)$ of the phase space where the numerically computed
first Lyapunov exponent $\lambda_1$ is very small goes to zero in a
rather fast way as $N$ increases
\begin{align}
P(N,\epsilon) \sim e^{-a(\varepsilon) N} \, .
\end{align}
At a first glance such a result sounds quite positive for the
possibility to build the statistical mechanics on dynamical bases,
since, no matter how small $\varepsilon$ is, the probability to end up
in a non-chaotic region becomes exponentially small in the size of the
system. At this stage we could say that one can consider himself quite
satisfied with the problem of dynamical justification of statistical
mechanics. KAM theorem told us that even for non integrable systems
there is a set of invariant manifolds of finite measure which
survives, and which may possibly spoil relaxation to equilibrium, but
very reasonable estimates also tell us that in the large-$N$ limit,
the probability to end up in a non-chaotic region is negligible. But,
as often happens in life, things are more complicated than they
seem. The problem is that the presence of chaos is not sufficient to
guarantee that some slow time-scales do not survive in the
system. From just an analysis in terms of Lyapunov exponents it is in
fact possible to conclude that \emph{almost} all trajectories have the
\emph{``good''} statistical properties we expect in statistical
mechanics.  The main reason is that the time
$\tau_{\text{Lyap}}=1/\lambda_1$, which is related to the trajectory
instability, is not the unique relevant characteristic time of the
dynamics. There might be more complicated collective mechanisms which
prevent a fast thermalization of the system which are not traced by
the Lyapunov spectrum. Think for instance to the mechanism of
ergodicity breaking in systems such as spin
glasses~\cite{book-MPV87}.\\

It is clear that as soon as $N \ge 3$ the motion lives in dimension
$2N-1$ while the KAM tori have dimension $N$, so that they are not
able to separate the phase space in disjoint components and, as we
have said, ergodicity is guaranteed asymptotically from the point of
view of dynamics. A trajectory initially closed to a KAM torus may
visit any region of the energy hypersurface, \emph{but}, since the
compenetration of chaotic regions and invariant manifolds typically
follows the pattern of a fractal geometry, the diffusion across the
isoenergetic hypersurface is usually very slow; as we said, such a
phenomenon is called Arnol'd diffusion~\cite{KB85,YK20}. The important and difficult
problem is therefore to understand the ``speed'' of the Arnold diffusion,
i.e. the time behaviour of
\begin{align}
\langle |\mI(t) -\mI(0) |^2 \rangle= \sum_{n=1}^N \langle |\mI_n(t) -\mI_n(0) |^2 \rangle
\end{align}
where the $\langle \,\,\, \rangle$ denotes the averages over the
initial conditions. There are some theoretical bounds for $ \langle
|\mI(t) -\mI(0) |^2 \rangle$, as well as several accurate numerical
simulations, which suggest a rather slow ``anomalous diffusion'':
\begin{align}
\langle |\mI(t) -\mI(0) |^2 \rangle \sim D \,  t^{ \nu} \,\,\, , \,\,  \nu  \le 1
\end{align}
where $D$ and $\nu$ depend from the parameters in the system. We can
say that in a generic Hamiltonian system Arnold diffusion is
present and, for small $\varepsilon$, it is very weak. It can thus
happen that different trajectories, even with a rather large Lyapunov
exponent, maintain memory of their initial conditions for considerable
long times. See for instance the results of~\cite{FMV91} discussed in
the Sec.~\ref{sec:s2} of this communication. The existence of Arnold's
diffusion is on the hints that chaos is sometimes not sufficient to
guarantee thermalization.\\ \\ 

\subsection{Lyapunov exponents in the large-$N$ limit}

Although we are going to depart from this point of view, let us put on
the table all evidences that chaos is apparently a good property to
justify a statistical mechanics approach. For instance in a chaotic
system it is possible to define an entropy directly from the Lyapunov
spectrum~\cite{book-PP16}. In any Hamiltonian (symplectic) system with $N$ degrees of
freedom we have $2N$ Lyapunov exponents $\lambda_1 \ge \lambda_2\ge
... \lambda_{2N -1} \ge \lambda_{2N}$ which obey a mirror rule,
i.e.
\begin{align}
\lambda_{2N} = -   \lambda_1  \,\, ,\,\,
\lambda_{2N-1} = -   \lambda_2  \,\, ,~~~\ldots
\end{align}
and the Kolmogorov-Sinai entropy is
\begin{align}
H=\sum_{n=1}^N \lambda_n \,.
\end{align}
If the interest is for the statistical mechanics, the natural question
is the asymptotic feature of $\{ \lambda_n \}$ for $N \gg 1$.  We have
clear numerical evidence~\cite{KK87,LPR86}, as well as a few
analytical results for special cases~\cite{BS93}, that at large $N$
one has
\begin{align}
\lambda_n= \lambda^* f(n /N)
\end{align}
where $\lambda^*=\lim_{N \to \infty} \lambda_1$. Such a result  implies that  the  Kolmogorov-Sinai entropy is
proportional to the number of degrees of freedom:
$$
 H \simeq h^* N
$$ where $h^*= \lambda^* \int_0^1 f(x) dx$ does not depend on $N$. The
 validity of such a thermodynamic limit is surely a positive result
 from the point of view of the statistical mechanics but it should not
 be overestimated, as we will discuss in the next section.

 \section{Challenging the role of chaos}
 \label{sec:s2}
 
 The previous section was entirely dedicated to present evidences in
 favour of chaos being a sufficient condition for an equilibrium
 behaviour in the large-$N$ limit and the possibility to apply a
 statistical mechanics description. In the present section we present
 counter-examples to this point of view. These examples suggest that
 in the large-$N$ limit a statistical mechanics description holds for
 almost all Hamiltonian systems irrespectively to the presence of
 chaos. First, we will quote two results showing that, even in the
 presence of finite Lyapunov exponents there are clear signatures of
 (weak) ergodicity breaking and of the impossibility to establish a
 statistical mechanics description on finite
 timescales~\cite{FMV91,LPRV87}. Conversely, we can also present the
 examples of an integrable system, the Toda chain~\cite{H74}, which
 shows thermalization on short time-scales notwithstanding all
 Lyapunov exponents are zero. Clearly in the case of an integrable
 system thermalization takes place with respect to some observables
 and not with respect to others. But this is the point of the whole
 discussion we want to promote: thermal behaviour is not a matter of
 the dynamics being regular or chaotic, it is just a matter of choosing
 the description of the system in term of the appropriate
 variables. And, as we will see that it is also the case for quantum
 mechanics (which has important similarities with classical integrable
 systems), almost all choices of canonical variables are good while
 only very few are not good. In the case of an integrable system in
 order to detect thermalization the canonical variables which
 diagonalize the Hamiltonian must be, of course, avoided.\\ \\ \\

\subsection{Weak ergodicity breaking in coupled symplectic maps}

 \begin{figure}
 \centering
   \includegraphics[width=0.9\columnwidth]{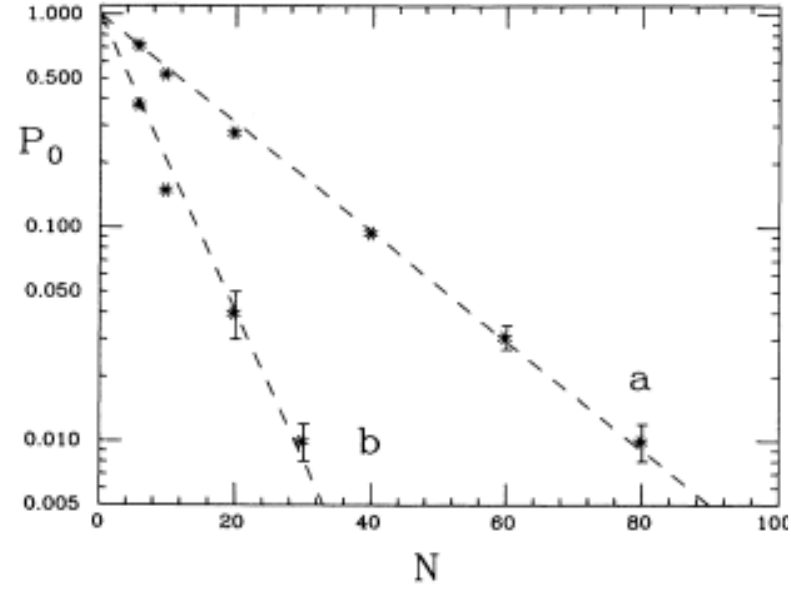}
   \caption{ Fraction of maximal Lyapunov exponents smaller than
     $0.002$ as a function of $N$ for (a) $\varepsilon=0.025$ and (b)
     $\varepsilon=0.05$, $T=10^5$, averaged over $\mN=500$
      initial conditions.}
\label{fig1}
 \end{figure}

We start our list of~\emph{``counter-examples''} to the idea that chaotic 
properties guarantee fast thermalization by recalling the results 
of~\cite{FMV91}. In~\cite{FMV91} the authors studied the ergodic properties of a 
large number of coupled symplectic maps~\eqref{eq:ham}, \eqref{eq:hamper} where  
one considers periodic boundary conditions, i.e.
$\theta_{N+1}=\theta_1,\, , \, I_{N+1}=I_1$,
  and nearest- neighbour coupling:
$$
F(\theta)=\sum_{n=1}^N \cos (\theta_{n+1}-\theta_n)\,.
$$

 As anticipated above, they find that the
 volume of phase space filled with invariant manifolds decreases
 exponentially with system size, as shown in Fig.~\ref{fig1}. This is
 a remarkable evidence that in the large-$N$ limit the probability
 that thermalization is obstructed by KAM tori is exponentially (in
 system size) small. The system has good chaotic properties. This
 notwithstanding, at the same time there are clear signatures of a
 phenomenon analogous to the so-called \emph{weak-ergodicity breaking}
 of disordered systems. The terminology weak-ergodicity breaking is a
 jargon indicating the situations where, despite relaxation
 is always achieved asymptotically, at any \emph{finite} time memory
 of the initial conditions is preserved~\cite{CK95}. Let us define as
 $t_w$ the time elapsed since the beginning of the dynamics,
 $t_w+\tau$ is a later time and $C(t_w,t_w+\tau)$ is the time
 auto-correlation (averaged over an ensemble of initial conditions) of
 some relevant observable of the system. Weak ergodicity breaking is
 then expressed by the condition
 \begin{align}
   & \lim_{\tau\rightarrow\infty} C(t_w,t_w+\tau) = 0 \quad \quad \quad \quad \quad
   \forall~~t_w\nonumber \\
   & C(t_w,t_w+\tau) \neq C(\tilde{t}_w,\tilde{t}_w+\tau) \quad \quad
   \forall~~\tau<\infty~~\&~~t_w\neq \tilde{t}_w
 \end{align}

 A clear signature of a situation with the features of weak ergodicity
 breaking is revealed by the study of how the distribution of the
 maximum Lyapunov spectrum depends on the system
 size~\cite{FMV91}. While the probability distribution of the
 exponents peaks at a well defined value at increasing system size,
 still the scaling of the variance is anomalous, i.e., it has a decay
 much slower than $1/\sqrt{N}$. One of the main results
 of~\cite{FMV91} is that the variance of the Lyapunov exponents
 distribution, in particular its asymptotic estimate
 $\sigma_\infty(\varepsilon,N)$ scales as
\begin{equation}
 \sigma_{\infty}(\epsilon, N) \sim  \frac{1}{N^{a(\epsilon)}}
\end{equation}
where $a(\epsilon) \sim \sqrt{\epsilon}$, so that for small
$\epsilon$,  $\sigma_{\infty}(\epsilon, N)$
is much larger that $1/\sqrt{N}$.

\subsection{High-temperature features in coupled rotators}

\begin{figure}
 \centering
   \includegraphics[width=0.9\columnwidth]{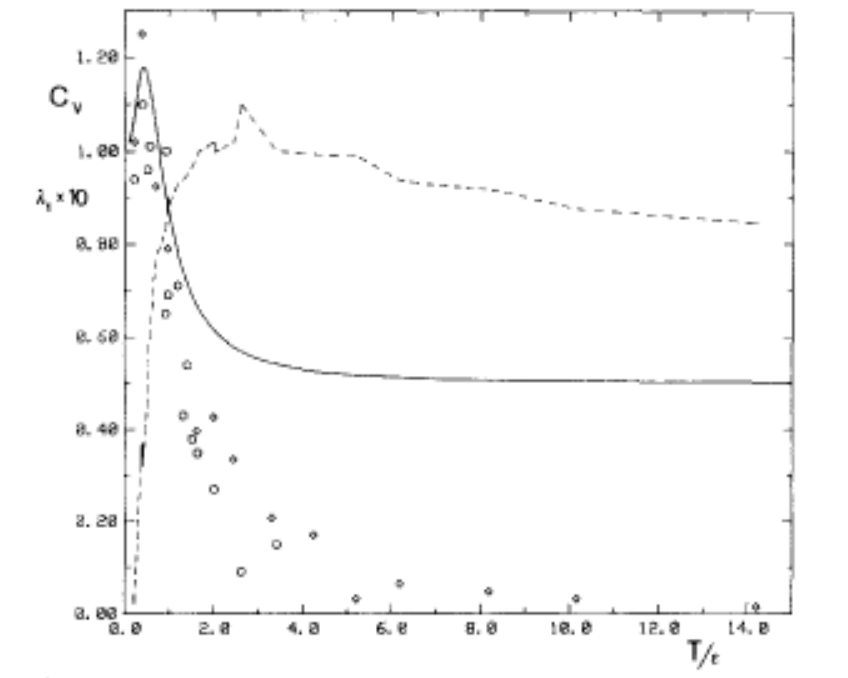}
   \caption{Specific heat versus temperature in the rotator model. The
     continuous line represents the analytic prediction from
     Eq.~(\ref{eq:CV-rot}). Empty symbols represents the results of
     numerical simulations, respectively for a subsystem with $M=10$
     (circles) and $M=20$ (diamonds) rotators, in a chain of $N=100$ and $N=400$ particles, respectively. The dashed line is the maximal
     Lyapunov exponent as a function of the temperature (specific
     energy) measured in units of the coupling constant $\varepsilon$
     [see Eq.~(\ref{eq:H-rotatori})].}
   \label{fig3}
 \end{figure}

 Another example of a system which does not show thermal behaviour
 despite having positive Lyapunov exponents is represented by the
 coupled rotators at high energy studied in~\cite{LPRV87}. The
 Hamiltonian of this system reads as:
\begin{align}
  \mH(\boldsymbol{\varphi},\boldsymbol{\pi}) = \sum_{i=1}^N \frac{\pi_i^2}{2} +
  \varepsilon~\sum_{i=1}^N (1-\cos(\varphi_{i+1}-\varphi_i)),
  \label{eq:H-rotatori}
\end{align}
where $\varphi_i$ are angular variables and $\pi_i$ are their
conjugate momenta. The specific heat can be easily computed from the
partition function as
\begin{align}
   C_V &= \frac{\beta^2}{N}~\frac{\partial^2}{\partial\beta^2}\log\mZ_N(\beta).
  \label{eq:Z-Cv-rot}
\end{align}
In particular, the expression of the specific heat reads in term of
modified Bessel function as
\begin{align}
  C_V = \frac{1}{2}+ \beta^2
  \left( 1 - \frac{1}{\beta} \frac{I_1(\varepsilon\beta)}{I_0(\varepsilon\beta)}-
  \left[ \frac{I_1(\varepsilon\beta)}{I_0(\varepsilon\beta)} \right]^2\right).
  \label{eq:CV-rot}
\end{align}
It is possible to compare the analytical prediction of
Eq.~(\ref{eq:CV-rot}) with numerical simulations by estimating in the
latter the specific heat from the fluctuations of energy in a given subsystem
of $M$ rotators, with $1 \ll M \ll N$:
\begin{align}
  C_V = \frac{1}{MT^2} \left[ \langle \mH_M ^2 \rangle - \langle \mH_M \rangle^2 \right]
  \label{eq:CV-num}
\end{align}
where $\mH_M$ is the Hamiltonian of the chosen subsystem and the ``ensemble'' average in Eq.~(\ref{eq:CV-num}) includes
averaging over initial conditions and averaging along the symplectic
dynamics for each initial condition. The result of the comparison is
presented in Fig.~\ref{fig3}. From the figure it is clear that, while
at small and intermediate energies the canonical ensemble prediction
for the specific heat matches numerical simulations, it fails at high
energies. What is remarkable is that in the high-temperature regime
where statistical mechanics fails the value of the largest Lyapunov
exponent is even larger that in the intermediate-small energy regime
where the equilibrium prediction works well. This is one of the
strongest hints from the past literature that the chaoticity of orbits
has nothing to do with the foundations of statistical mechanics. In
practice it happens that for rotators a sort of ``effectively
integrable'' regime arises at high energy. In this regime each rotator
is spinning very fast, something which guarantees the chaoticity of
orbits, but the individual degrees of freedom do not interact each
other, which causes the breakdown of thermal properties of the
system. This very simple mechanism is a concrete example (and often
examples are more convincing than arguments) of how absence of
thermalization and chaos can be simultaneously present without any
problem. With the next example the role of chaos for thermalization
will be challenged even further: we are going to present the case of a
system which, despite having \emph{all Lyapunov exponents equal to
  zero}, relaxes nicely to thermal equilibrium on short
time-scales.\\ \\

\subsection{Toda lattice: thermalization of an integrable
     system}
\label{sub:Toda}
 
 We present here the example of an integrable system which shows very
 good thermalization properties.  In some sense we find that the
 statement ``the system has thermalized'' or ``the systems has not
 thermalized'' depends, for Hamiltonian systems, on the choice of
 canonical coordinates in the same way as the statement ``a body is
 moving'' or ``a body is at rest'' depend on the choice of a reference
 frame. In this respect, it is true that, if a system is integrable
 in the sense of the Liouville-Arnol'd theorem and we choose to
 represent it in terms of the corresponding action-angle variables, we
 will never observe thermalization. But there are infinitely many
 other choices of canonical coordinates which allow one to detect a
 good degree of thermalization. Let us be more specific about this and
 recall the salient results presented
 in~\cite{BVG20} on the Toda chain. 
 
 \begin{figure}
  \centering
  \includegraphics[width=0.9\linewidth]{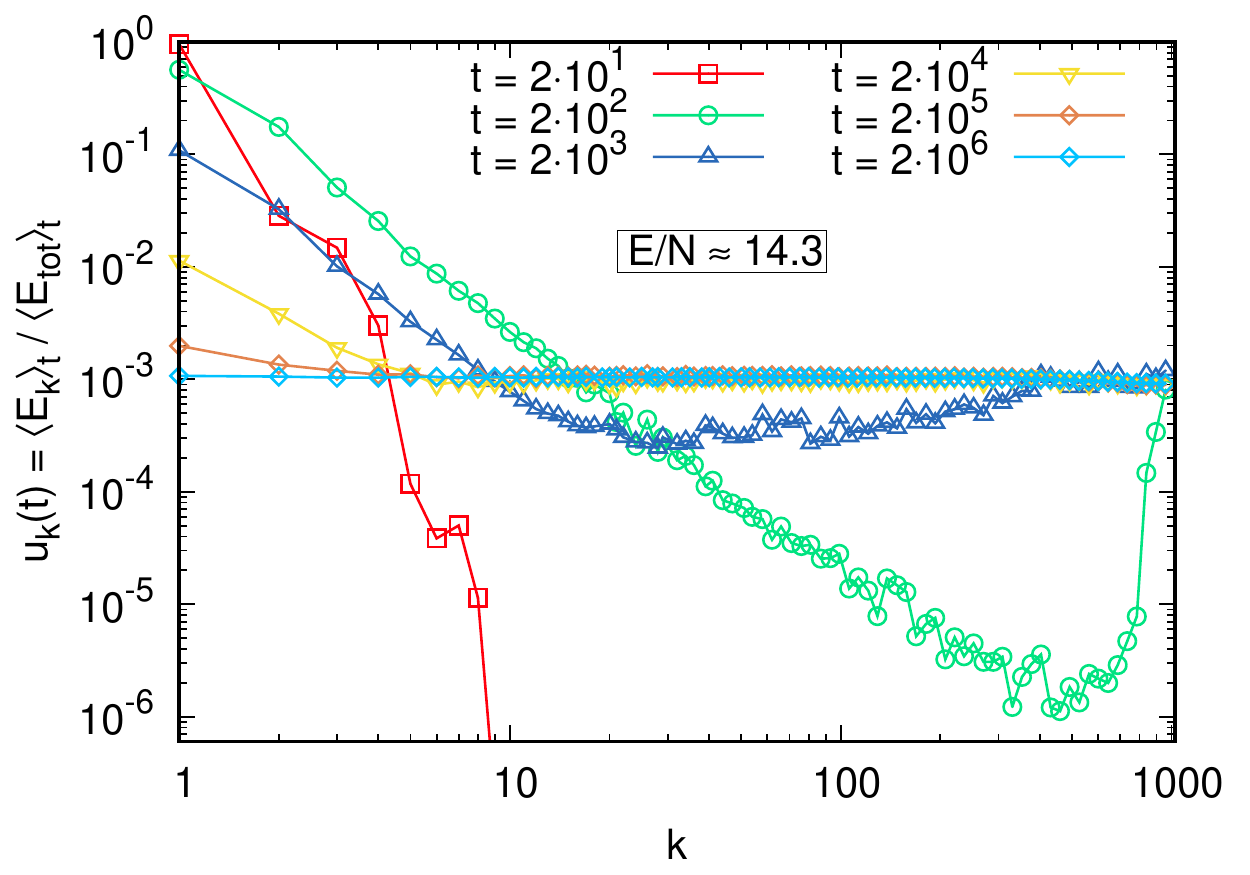}
  \caption{\label{fig:toda} Normalized harmonic energy as a function of the
  Fourier mode number $k$, for different values of the averaging time, in a Toda dynamics~\eqref{eq:HToda}
  with FPUT initial conditions. Here $N=1023$, total energy $E=14.3$.}
 \end{figure}

 It is well known that the Toda lattice 
\begin{align}
  \mH(q,p) = \sum_{i=1}^N \frac{p_i^2}{2} + \sum_{i=0}^N V(q_{i+1}-q_i),
  \label{eq:HToda}
\end{align}
where $V(x)$ is the Toda potential
\begin{align}
  V(x) = \exp(-x)+x-1,
  \label{eq:Vx}
\end{align}
admits a complete set of independent integrals of motion, as shown by Henon in 
his paper~\cite{H74}. This result can be also understood in the light of 
Flashka's proof of the existence of a Lax pair related to the Toda 
dynamics~\cite{F74}. The explicit form of such integrals of motion is rather 
involved, and their physical meaning is not transparent; one may wonder whether 
the system is able to reach thermalization under a different canonical 
description, for instance the one which is provided by the Fourier modes
 \begin{align}
  Q_k &= \sum_{i=1}^N\frac{\sqrt{2}~q_i}{\sqrt{N+1}}\sin\left( \frac{\pi i k}{N+1}\right)\, \nonumber \\
  P_k &= \sum_{i=1}^N\frac{\sqrt{2}~p_i}{\sqrt{N+1}}\sin\left( \frac{\pi i k}{N+1}\right)\,.
\end{align}
 
A typical test which can be performed, inspired by the FPUT numerical
experiment, is realized by considering an initial condition in which
the first Fourier mode $k=1$ is excited, and $Q_k=P_k=0$ for all
$k>1$; the system is evolved according to its Hamiltonian dynamics,
and the average energy corresponding to each normal mode,
\begin{align}
  \langle E_k\rangle_t &= \int_0^t \left[\frac{P_k^2}{2}+ \omega_k^2 \frac{Q_k^2}{2}\right]\, dt\, \nonumber \\
  \omega_k&=2 \sin \left(\frac{\pi k}{2 (N+1)}\right)\,,
\end{align} 
is studied as a function of time. It is worth noticing that in the
limit of small energy $E$, the Toda Hamiltonian is well approximated
by a harmonic chain in which the $P_k$'s are conserved quantities; it
is therefore not surprising that, for $E/N \ll 1$, it takes very long
times to observe equipartition of harmonic energy between the
different modes. This ergodicity-breaking phenomenology has been
studied, for instance, in Ref.~\cite{BCP13} to underline relevant
similarities with the FPUT dynamics.

The scenario completely changes when specific energies of order
$\approx \mathcal{O}(1)$ or larger are considered.  In
Fig.~\ref{fig:toda} the distribution of the harmonic energy among the
different Fourier modes is shown for several values of the averaging
time $t$; no ergodicity breaking can be observed, and in finite times
the harmonic energy reaches equipartition.


 \section{From classical to quantum: open research directions}
 \label{sec:s3}

 After the examples of the previous section we feel urged at this
 point to draw some conclusions. We started our discussion from
 wondering how much having a phase space uniformly filled by chaotic
 regions is relevant for relaxation to equilibrium and we came to the
 conclusion, drawn from the last set of examples, that in practice
 chaos seems to be at all irrelevant for relaxation to equilibrium.
 What it is then the correct way to frame the ``foundation of
 statistical mechanics'' problem? It seems that the correct paradigm
 to understand the results of Sec.~\ref{sec:s2} is the one established
 by the ergodic theorem by Khinchin~\cite{book-K49}: in order to
 guarantee a thermal behaviour, i.e., that dynamical averages
 correspond to ensemble averages, it is sufficient to consider
 appropriate observables for large enough systems. That is, the regime
 $N \gg 1$ is mandatory. As anticipated in Sec.~\ref{sec:s2} by our
 numerical results, one finds that the hypothesis under which the
 Khinchin ergodic theorem is valid are for instance fulfilled even by
 integrable systems. In a nutshell we can summarize his approach
 saying that it is possible to show the practical validity of the
 ergodic hypothesis when the following three conditions are fulfilled:
\begin{itemize}
\item[(a)] the number of the degrees of freedom is very large
\item[(b)] we limit the interest to ``suitable'' observables
\item[(c)] we allow for a failure of the equivalence of the time
  average and the ensemble average for initial conditions in a
  ``small region''.
\end{itemize}
The above points will be clarified in the following.

\subsection{The ergodic problem in Khinchin's perspective}

 Let us see in practice what the theorem says. Consider a Hamiltonian
 system $\mH({\bf q},{\bf p})$ with canonical variables $({\bf q},{\bf
   p}) = (q_1,p_1,\ldots,q_N,p_N)$. Let $\Phi_t({\bf q},{\bf p})=({\bf
   q}(t),{\bf p}(t))$ be the flow under the symplectic dynamics
 generated by $\mH({\bf q},{\bf p})$. The time average for given
 initial datum $x=({\bf q},{\bf p})$ is defined as
 \begin{align}
   \overline{f(x)} = \frac{1}{T} \int_0^T dt~f(\Phi_t(x))  
   \label{eq:time-av}
 \end{align}
 while the ensemble average is defined with respect to an invariant
 measure $\mu({\bf q},{\bf p})$ in phase space:
 \begin{align}
   \langle f \rangle = \int~d\mu({\bf q},{\bf p})~f({\bf q},{\bf p}).
   \label{eq:phase-av}
 \end{align}

 Let us define as \emph{sum function} any function reading as
 \begin{align}
   f(x)=\sum_{i=1}^N f_i({\bf q}_i, {\bf p}_i)
 \end{align}

 For such functions Khinchin~\cite{book-K49} was able to show that for a random choice
 of the initial datum $x$ the probability that the time average in
 Eq.~(\ref{eq:time-av}) and the ensemble average in
 Eq.~(\ref{eq:phase-av}) are different is small in $N$, that is:
 \begin{align}
   p\left( \left|\frac{\overline{f(x)} - \langle f \rangle}{\langle f \rangle}\right| \geq \frac{C_1}{N^{1/4}} \right)
   < \frac{C_2}{N^{1/4}},
   \label{eq:erg-khinchin}
 \end{align}
 where $C_1$ and $C_2$ are constants with respect to $N$.  We can note
 that many, but not all, relevant macroscopic observables are sum
 functions. While the original result of Khinchin was for non
 interacting system, i.e. systems with Hamiltonian
\begin{align}
 \mH({\bf q},{\bf p})=\sum_{i=1}^N h_1(q_i, p_i)
\end{align}
Mazur and van der Linden~\cite{ML63} were able to generalize it to the more
physical interesting case of (weakly) interacting particles:
\begin{align}
 H=\sum_{i=1}^N h_1(q_i, p_i)+ \sum_{i<j} V(|q_i - q_j|)
\end{align}
From the results by Khinchin, Mazur and van der Linden we have the
following scenario: although the ergodic hypothesis mathematically
does not hold, it is ``physically'' valid if we are tolerant, namely
if we accept that in systems with $N\gg1$ ergodicity can fail in
regions sampled with probability of order $O(N^{-1/4})$, i.e.
vanishing in the limit $N \to \infty$.  Let us stress that the
dynamics has a marginal role, while the very relevant ingredient is
the large number of particles.\\

Whereas it is possible, on the one hand, to emphasize some limitations
of the Khinchin ergodic theorem, as for instance the fact that it does
not tell anything about the timescale to be waited for in order to
have Eq.~(\ref{eq:erg-khinchin}) reasonably true, let us stress here
its goals. For instance, let us highlight the fact that Khinchin
ergodic theorem guarantees the thermalization even of \emph{integrable}
systems! In fact, if we consider the Toda model discussed in the
previous section, the conditions under which the Khinchin ergodic
theorem was first demonstrated perfectly apply to it.  In fact
integrability, proved first by H\'enon in 1974~\cite{H74}, guarantees
the existence of $N$ action-angle variables such that the Hamiltonian
reads as:
\begin{align}
\mH(\boldsymbol{\mI},\boldsymbol{\phi})=\sum_{i=1}^N h_1(\mI_i, \phi_i),
\end{align}
which is precisely the \emph{non-interacting-type} of Hamiltonian
considered in first instance by Khinchin. The results presented
in~\cite{BVG20} on the fast thermalization of the Toda model has been
in fact proposed to re-establish with the support of numerical
evidence the assertion, hidden between the lines of the Khinchin
theorem and perhaps overlooked so far, that even integrable systems do
thermalize in the large-$N$ limit. The phase-space of Toda chain is in
fact completely foliated in invariant tori. But, according to Khinchin
theorem and our numerical results, this foliation of phase space in
regular regions is not an obstacle as long as the
\emph{``ergodization''} of sum functions is considered (for almost all
initial data in the limit $N\rightarrow\infty$).\\
\subsection{Quantum counterpart: Von Neumann's ergodic theorem}
The observation that \emph{``even integrable systems thermalize well
  in the large-$N$ limit''} becomes particularly relevant as soon as
we regard, in the limit of large $N$, the dynamics of a classical
integrable system as the \emph{``classical analog''} of the dynamics
of a quantum system.  Indeed in quantum system arises the same
``ergodic problem'' that we find in classical mechanics, i.e., whether
time averages can be replaced by ensemble averages. Let us spend few
words on the quantum formalism in order to point out the similarities
between the ergodic theorem by Khinchin and the one by Johan Von
Neumann for quantum mechanics. As it is well known, due to the
self-adjointness of the Hamiltonian operator, any wave vector
$|\psi\rangle$ can be expanded on the Hamiltonian eigenvectors basis:
\begin{align}
|\psi\rangle = \sum_{\alpha \in \text{Sp}(\hat{H})} c_\alpha~| \alpha \rangle
\end{align}
The projection on a limited set of eigenvalues defines the quantum
microcanonical ensemble. Since it is reasonable to assume that also in
a quantum system the total energy of an $N$-particles system is known
with finite precision, i.e., usually we know that it takes values
within a finite shell $\mI_E \in [E-\delta E , E+\delta E]$ with
$E\sim N$ and $\delta E \sim \sqrt{N}$, one defines the microcanonical
density matrix as the projector on the eigenstates pertaining to that shell:
\begin{align}
\hat{\rho}_E = \frac{1}{\mN_E} \sum_{\alpha | \varepsilon_\alpha \in \mI_E} | \alpha \rangle~\langle \alpha |,
\end{align}
where $\mN_E$ is the number of eigenvalues in the shell. The microcanonical
expectation of the observable thus read
\begin{align}
  \langle \hat{O}\rangle_E ~=~ \text{Tr}[\hat{\rho}_E \hat{O}] ~=~ 
  \frac{1}{\mN_E} \sum_{\alpha | \varepsilon_\alpha \in \mI_E} \langle \alpha | \hat{O}| \alpha \rangle
\end{align}
Clearly in
the limit $N\rightarrow\infty$ where the eigenvalues tend to fill
densely the real line one has that even in a finite shell $\mN_E \sim
N$. By preparing a system in the initial state
$|\psi_0\rangle$ the expectation value of a given
observable $\hat{O}$ at time $t$ reads as:
\begin{align}
  \langle \hat{O}(t) \rangle_{\psi_0} = \langle \psi_0 |~e^{i \hat{H}t/\hbar}~\hat{O}~e^{-i \hat{H}t/\hbar}~| \psi_0 \rangle
  = \langle \psi_0(t) | \hat{O} | \psi_0(t) \rangle \nonumber \\
\end{align}
\gcol{ \emph{Quantum ergodicity} amounts then to the following
  equivalence between dynamical and ensemble averages,
\begin{align}
\langle \hat{O}\rangle_E ~=~\lim_{T\rightarrow\infty} \frac{1}{T} \int_0^T dt~\langle \hat{O}(t) \rangle_{\psi_0}.
\label{eq:quantum-erg}
\end{align}
The reader can easily convince himself of the fact that, if for
\emph{almost all} times the expectation of $\hat{O}(t)$ on the initial
state $|\psi_0\rangle$ is \emph{typical}, namely one has $\langle
\hat{O}\rangle_E \approx \langle \hat{O}(t) \rangle_{\psi_0}$, then
the quantum ergodicity property as stated in
Eq.~(\ref{eq:quantum-erg}) is realized. Having for almost all times
$\langle \hat{O}\rangle_E \approx \langle \hat{O}(t) \rangle_{\psi_0}$
is a condition named \emph{``normal typicality''}: it is discussed
thoroughly in~\cite{GLMTZ10}.
Remarkambly in tune with the scenario later proposed by Kinchin for
classical system, already in 1929 Von Neumann proposed a quantum
ergodic theorem which proves the \emph{typicality} of $\langle
\hat{O}(t) \rangle$ with no references to \emph{quantum chaos}
properties. The definition of quantum chaos, which will be formalized
much later~\cite{BGS84}, is to have a system characterized by a
Hamiltonian operator, $\hat{H}|\alpha \rangle = \varepsilon_\alpha |
\alpha \rangle$, such that its eigenvctors $|\alpha\rangle$ behave as
random structureless vectors in any basis. This property,
notwidthstanding the different formalism of quantum and classical
mechanics, has a deep analogy with the definition of classical
chaos~\cite{casati2006,book-CCV09}: a quantum system is said to be
chaotic when a small perturbation of the Hamiltonian, $\hat{H} +
\hat{\lambda} $ produces totally uncorrelated vectors, $\langle
\epsilon_\alpha | \epsilon_\alpha+\lambda_\alpha\rangle \sim
1\/\sqrt{N}$, in the same manned that a small shift in initial
conditions produces totally uncorrelated trajectories in classical
chaotic systems.\\ Quite remarkably, Von Neumann's ergodic theorem
does not make any explicit assumption of the above kind on the
Hamiltonian's eigenvalues structure. Here follows a short account of
the theorem, mathematical details can be found in the recent
translation from German of the original paper~\cite{JVN10} and
in~\cite{GLMTZ10,GLTZ10}. Let $\mD$ be the dimensionality of the
energy shell $\mI_E = [E-\delta E, E+ \delta E]$, namely
$\mD=\text{dim}(\mH_E)$ where $\mH_E$ is the Hilbert space spanned by
the eigenvectors $|\alpha\rangle$ such that $\varepsilon_\alpha \in
\mI_E$, and define a decomposition of $\mH_E$ in orthogonal subspaces
$\mH_\nu$ each of dimension $d_\nu$:
\begin{align}
\mH_E = \bigoplus_{\nu} \mH_\nu \quad\quad\quad \text{dim}(\mH_\nu)=d_\nu \quad\quad\quad \sum_\nu d_\nu = \mD
\end{align}
Then define $\hat{P}_\nu$ as the projector on subspace $\mH_\nu$. It
is mandatory to consider the large system size limit where $1\ll d_\nu
\ll \mD$. It is then demonstrated that, under quite generic
assumptions on $\hat{H}$ and on the orthogonal decomposition $\mH_E =
\bigoplus_{\nu} \mH_\nu$, for \emph{every wavefunction} $|
\psi_0\rangle \in \mH_E$ and for \emph{almost all} times one has normal typicality, i.e.:
\begin{align}
  \langle \psi_0 | \hat{P}_\nu(t)| \psi_0 \rangle = \text[\hat{\rho}_E \hat{P}_\nu].
\end{align}
A crucial role is played by the assumption that the dimensions $d_\nu$
of the orthogonal subspaces of $\mH_E$ are macroscopically large,
i.e. $d_\nu \gg 1$. In this sense the projectors $P_\nu$ correspond to
macroscopic observables. From this point of view the quantum ergodic
theorem from Von Neumann is constrained by the same key hypothesis of
the Khinchin's theorem: the limit of a very large number of degrees of
freedom. At the same time the Von Neumann theorem does not make any
claim on chaotic properties of the eigenspectrum, in the very same way
as the Khinchin's theorem does not make any claim on chaotic
properties of trajectories.\\ For completeness we also need to
mention what is regarded nowday as the ``modern'' version of the Von
Neumann theorem, namely the celebrated Eigenstate Thermalization
Hypothesis (ETH)~\cite{RS12}. By expanding the expression of $\langle
\hat{O}(t) \rangle_{\psi_0}$ as
\begin{align}
  \langle \hat{O}(t) \rangle_{\psi_0} =
  \sum_{\alpha\in\text{Sp}\hat{H}} |c_{\alpha}|^2~O_{\alpha\alpha} +
  \sum_{\alpha\neq\beta} e^{i(E_\alpha-E_\beta)t/\hbar} ~c_{\alpha}^*
  c_\beta~O_{\alpha\beta}.
  \end{align}
it is not difficult to figure out that ergodicity, as expressed in
Eq.~(\ref{eq:quantum-erg}), is guaranteed if suitable hypotheses are
made for the matrix elements $O_{\alpha\beta}$. For instance it can be
assumed that:
\begin{itemize}
  \item[h.a:] The off-diagonal matrix elements are exponentially small
    in system size, $O_{\alpha\neq\beta} \sim e^{-S(E_0)}$ with $S$
    the microcanonical entropy and $E= \langle \psi_0 |
    \hat{H}|\psi_0\rangle$.

  \item[h.b:] The diagonal elements are a function of the
    initial-condition energy, $O_{\alpha\alpha} = O(E_0)$.
\end{itemize}
The hypothesis $h.a$ guarantees that \emph{for large enough systems}
relaxation to a stationary state is achieved within reasonable time,
still leaving open the possibility that stationarity is different from
thermal equilibrium. In fact, it is thanks to the exponentially small
size of non-diagonal matrix elements that in the large-$N$ limit one
does not need to wait the astronomical times needed for dephasing in
order to have
\begin{align}
\frac{1}{T}\int_0^T dt~\sum_{\alpha\neq\beta} e^{i(E_\alpha-E_\beta)t/\hbar} ~c_{\alpha}^*
c_\beta~O_{\alpha\beta} ~\approx~ 0
\label{eq:dephasing}
\end{align}
Hypothesis $h.b$ then guarantees that relaxation is towards a state
well characterized macroscopically, i.e. a state which depends solely
on the energy of the Hilbert space vector $|\psi_0\rangle$ and not on
the extensive number of coefficients $c_\alpha$:
\begin{align}
\sum_{\alpha\in\text{Sp}\hat{H}} |c_{\alpha}|^2~O_{\alpha\alpha} =
O(E_0)\cdot \sum_{\alpha\in\text{Sp}\hat{H}} |c_{\alpha}|^2 = O(E_0)
\end{align}
Though inspired from the behaviour of (quantum) chaotic systems, ETH
is clearly a \emph{different} property since it makes no claim on the
structure of energy eigenvectors. For this reason and also because a
necessary condition for ETH to be effective is a large-$N$ limit, ETH
is rather close in spirit to the Von Neumann theorem. Then, we must
say that a complete understanding of the reciprocal implications of
quantum chaos and the Eigenstate Thermalization Hypothesis is still an
open issue which deserves further investigations. To this respect let
us recall the purpose of this chapter, namely to furnish reasons to
believe that an investigation aimed at challenging the role of chaos
in the thermalization of both quantum and classical systems is an
interesting and timely subject. For instance, a deeper understanding
of the mechanisms which triggers, or prevents, thermalization in
quantum systems is crucial to estimate the possibilities of having
working scalable quantum technologies, like quantum computers and
quantum sensors.\\

An interesting point of view which emerged from the results on the
Toda model discussed in Sec.~\ref{sub:Toda} and which can be traced
back to both Khinchin and Von Neumann's theorem is the following: the
presence (or absence) of thermalization is a property pertaining to a
given choice of observables and cannot be stated in general,
i.e. solely on the basis of the behaviour of trajectories or the
structure of energy eigenvectors. We have in mind the choice of the
projectors $P_\nu$ in Von Neumann theorem and the choice of sum
functions in Khinchin theorem. This perspective of considering
``thermal equilibrium as a matter of observables choice'' is in our
opinion a point of view which deserves a careful investigation.}

\gcol{
  
\subsection{Summary and Perspectives}

Let us try to summarize the main aspects here discussed.  At first we
stress that the ergodic approach, even with some caveats, appears the
natural way to use probability in a deterministic context.  Assuming
ergodicity, it is possible to obtain an empirical notion of
probability which is an objective property of the trajectory. An
important aspect often non considered is that both in experiments and
numerical computation one deals with a unique system with many degrees
of freedom, and not with an ensemble of systems.  According to the
point of view of Boltzmann (and the developments by Khinchin, Mazur and
van der Linden) it is rather natural to conclude that, at the
conceptual level, the only physically consistent way to accumulate a
statistics is in terms of time averages following the time evolution
of the system. At the same time ergodicity is a very demanding
property and, since in its definition it requires to consider the
infinite time limit, physically it is not very accessible.\\

We have then presented a strong evince from numerical study of high
dimensional Hamiltonian systems that chaos is neither a necessary nor
a sufficient ingredient to guarantee the validity of equilibrium
statistical mechanics for classical systems. On the other hand, even
when chaos is very weak (or absent), we have shown examples of a good
agreement between time and ensemble averages~\cite{BVG20}.  The
perspective emerging from our study is that the choice of variables is
thus a key point to say whether a system has thermalized or not. We
have found that this point of view emphasizes the commonalities
between classical and quantum mechanics for what concerns the ergodic
problem. This is particularly clear by comparing the assumptions and
the conclusions of the ergodic theorems of Khinchin and Von Neumann,
where, without \emph{any} assumption on the chaotic nature of
dynamics, it is shown that for \emph{general enough observables} a
system has good ergodic properties even in the case where interactions
are absent, provided that the system is large enough.\\

This brought us to underline as a possibly relevant research line the
one dedicated to find ``classical analogs'' of thermalization problems
in quantum systems and to study such problems with the conceptual
tools and the numerical techniques developed for classical Hamiltonian
systems.\\

The ability to control and exactly predict the behavior of quantum
systems is in fact of extreme relevance, in particular for the great
bet presently made by the worldwide scientific community on quantum
computers. In particular, understanding the mechanisms preventing
thermalization might certainly help to improve the function of quantum
processors and to devise better quantum algorithms.\\

In summary, we tried to present conving motivations in favour of a
renewed interest towards the foundations of quantum and statistical
mechanics. This is something which in our opinion is worth
to be pursued while approaching the one century anniversary of the two
seminal papers of Heisenberg~\cite{He25} and
Schr\"odinger~\cite{Sch26} on quantum mechanics. How much the quantum
mechanics revolution has influenced across one hundred year not only
the understanding of the microscopic world but even the thermodynamic
properties of macroscopic systems?  This the, still actual, key
question of our reserch line.}

\begin{acknowledgements}
We thank for useful discussions R. Livi, V. Ros, A. Scardicchio and
N. Zangh\`i. M.B. and A.V. acknowledge partial financial support of
project MIUR-PRIN2017~\emph{``Coarse-grained description for
  non-equilibrium systems and transport phenomena''}
(CO-NEST). G.G. thanks the Physics Department of ``Sapienza'',
University of Rome, for kind hospitality during some stages of this
work preparation.
\end{acknowledgements}


\end{document}